\documentclass[aps,prl,superscriptaddress,twocolumn]{revtex4}

\newcommand{\bfk}{\mbox{\boldmath $k$}}
\newcommand{\bfx}{\mbox{\boldmath $x$}}
\newcommand{\bfy}{\mbox{\boldmath $y$}}

\begin{document}

\title{Comment on ``Quantum back-reaction through the Bohmian particle''}

\author{L.L. Salcedo}

\affiliation{
Departamento de F\'{\i}sica Moderna,
Universidad de Granada,
E-18071 Granada, Spain
}

\date{\today}



\maketitle

In a recent Letter \cite{prezhdo:2001none} Prezhdo and Brooksby have
forwarded a solution to the long-standing problem of properly dealing with
a mixed quantum-classical system, namely, to couple the classical
sector, not directly to the quantum one, but to its associated Bohmian
particle \cite{Holland:1993bk}. The precise proposal is to make
replicas of the quantum-classical system, each one with a different
position of the Bohmian particle obtained by sampling the ``initial''
quantum sector density. Each copy of the system is then to be evolved
as follows: i) the quantum sector evolves as usual, i.e., by taking
the classical degrees of freedom as time-dependent parameters of the
quantum Hamiltonian, ii) the Bohmian particle evolves {\em a la} de
Broglie-Bohm, its velocity depending solely on the quantum degrees of
freedom and iii) the classical sector evolves following Hamilton's
equations but this time with the Bohmian particle position as
time-dependent parameter in the Hamiltonian. The expectation values of
observables are obtained by averaging over the ``final'' sample of
quantum-classical degrees of freedom. In practice this proposal
coincides with that in \cite{gindensperger:2000none} (in its simplest
version) although there it was derived as an approximation rather than
introduced as a postulate.  In this Comment I want to point out some
limitations of the proposed prescription if it is pushed too far.

Without loss of generality, it can be assumed that the initial state
is given by some probability distribution $n(u,\psi)$ defined on the
combined manifold with coordinates $u=(\bfx,\bfk)$ (the
position-momentum of the classical particle) and $|\psi\rangle$ (the
normalized wave-function of the quantum particle). The full
distribution $f(u,\psi,\bfy)$ including the Bohmian particle (with
coordinates $\bfy$) will then equal $n(u,\psi)\,|\psi(\bfy)|^2$.  In
this way $f$ is initially fully determined by $n$. The evolution
equations then define a flow for $f$ in the manifold
$(u,\psi,\bfy)$. The flow of $f$ has no diffusion (an initial Dirac
delta distribution evolves as a delta all the time). Of course, its
marginal density $n(u,\psi;t)=\int d\bfy f(u,\psi,\bfy;t)$ does show
diffusion, which is interpreted as branching and back-reaction of the
quantum-classical system.

The first problem of the proposal is that the initial relation noted
between $f$ and $n$ is not preserved by the evolution, that is, at
later times the Bohmian particle is no longer distributed as
$|\psi(\bfy)|^2$ since a bias is introduced by its coupling to the
classical particle coordinates. Some negative consequences of this
fact are: i) The proof given in \cite{prezhdo:2001none} that the total
quantum-classical energy is conserved no longer applies. In fact
energy is not conserved (no such claim was made in
\cite{gindensperger:2000none}). ii) Unlike $f$, the distribution
$n(u,\psi;t)$ does not satisfy an evolution equation, that is, the
knowledge of this distribution at some intermediate time $t$ is
insufficient to predict its future evolution since the distribution of
the Bohmian particle is also needed. Unfortunately, physics lies in
$n$ not $f$: If the proposal in the Letter is taken seriously as a
prescription to evolve $n$ from an initial time $t$ to a final time
$t^\prime$ (let us denote this operation by $U(t^\prime,t)$) it will
be found that $U(t_2,t_1)U(t_1,t_0)$ differs from $U(t_2,t_0)$ since
the proposed prescription requires to re-sample the Bohmian particle
distribution at the intermediate time $t_1$.

A second problem is as follows. As is well known in pure quantum
mechanics, given an ensemble of pure states, only the combination
$\sum_i p_i|\psi_i\rangle\langle\psi_i|$ is relevant to physics, since
nothing else can be extracted from the knowledge of the expectation
values of all possible observables. For the same reason, in the mixed
quantum-classical case, only the quantity $\hat\rho(u)= \int
D\psi\,n(u,\psi)\,|\psi\rangle\langle\psi|$ (the mixed density
matrix-density distribution in phase-space) characterizes the physical
state. Note that there are many different $n$ producing the same
$\hat\rho$. The trouble is that in the proposed method, two different
initial distributions $n$ corresponding to the same $\hat\rho$ will in
general evolve into different $\hat\rho$ at later times (as it is
readily shown), in other words, identical initial physical states
$\hat\rho$ will branch into different states $\hat\rho(t)$ at later
times, depending on which initial distribution $n$ was chosen to
represent it. This is a general problem whenever the evolution is not
directly based on the quantity $\hat\rho(u)$.

This said, we have to distinguish between practical and theoretical
points of view. At a practical level the method has been found to be
useful and give good results in concrete applications
\cite{prezhdo:2001none,gindensperger:2000none,gindensperger:2002none}.
Our critique implies, however, that at an exigent theoretical level
the method fails to be fully internally consistent (this merely
reflecting its intrinsically approximate nature) and cannot be
regarded as a final answer to the quantum-classical mixing problem.

\begin{acknowledgments} This work is supported in part by funds provided
by the Spanish DGICYT grant no. PB98-1367 and Junta de Andaluc\'{\i}a
grant no. FQM0225.
\end{acknowledgments}


\end{document}